\documentclass[%
 reprint,
superscriptaddress,
amsmath,amssymb,
aps,
pra,
]{revtex4-1}
\usepackage{caption}
\usepackage{subcaption}
\usepackage{graphicx}
\usepackage{xcolor}
\usepackage{graphicx}
\usepackage{dcolumn}
\usepackage{bm}

\begin{document}

\preprint{APS/123-QED}

\title{Isochronous space-time wave packets}

\author{Alyssa A. Allende Motz}
\affiliation{CREOL, The College of Optics \& Photonics, University of Central Florida, Orlando, FL 32816, USA}
\author{Murat Yessenov}
\affiliation{CREOL, The College of Optics \& Photonics, University of Central Florida, Orlando, FL 32816, USA}
\author{Ayman F. Abouraddy}
\thanks{corresponding author: raddy@creol.ucf.edu}
\affiliation{CREOL, The College of Optics \& Photonics, University of Central Florida, Orlando, FL 32816, USA}




\begin{abstract}
The group delay incurred by an optical wave packet depends on its path length. Therefore, when a wave packet is obliquely incident on a planar homogeneous slab, the group delay upon traversing it inevitably increases with the angle of incidence. Here we confirm the existence of isochronous `space-time' (ST) wave packets: pulsed beams whose spatio-temporal structure enables them to traverse the layer with a fixed group delay over a wide span of incident angles. This unique behavior stems from the dependence of the group velocity of a refracted ST wave packet on its angle of incidence. Isochronous ST wave packets are observed in slabs of optical materials with indices in the range from 1.38 to 2.5 for angles up to $50^{\circ}$ away from normal incidence.
\end{abstract}

\maketitle

Fermat demonstrated in 1662 that a minimal principle underpins Snell's law; namely, the refracted trajectory corresponds to a minimum optical path length \cite{Sabra81Book}. Identifying minimal principles is now a commonplace strategy throughout the physical sciences. A different strategy is to search for isochronous (or equal-time) configurations. One example is Huygens' isochronous pendulum, which is built on the discovery that a particle sliding on a cycloid (or tautochrone curve) returns after a fixed time interval independently of point of release \cite{Huygens86Book,Randazzo18AJP}. However, this strategy has had less impact in optics. In general, the group delay accrued by a pulse in a homogeneous isotropic medium is proportional to the distance traveled, and no isochronous configurations arise except along equal-length paths.

\begin{figure}[t!]
  \begin{center}
  \includegraphics[width=8.6cm]{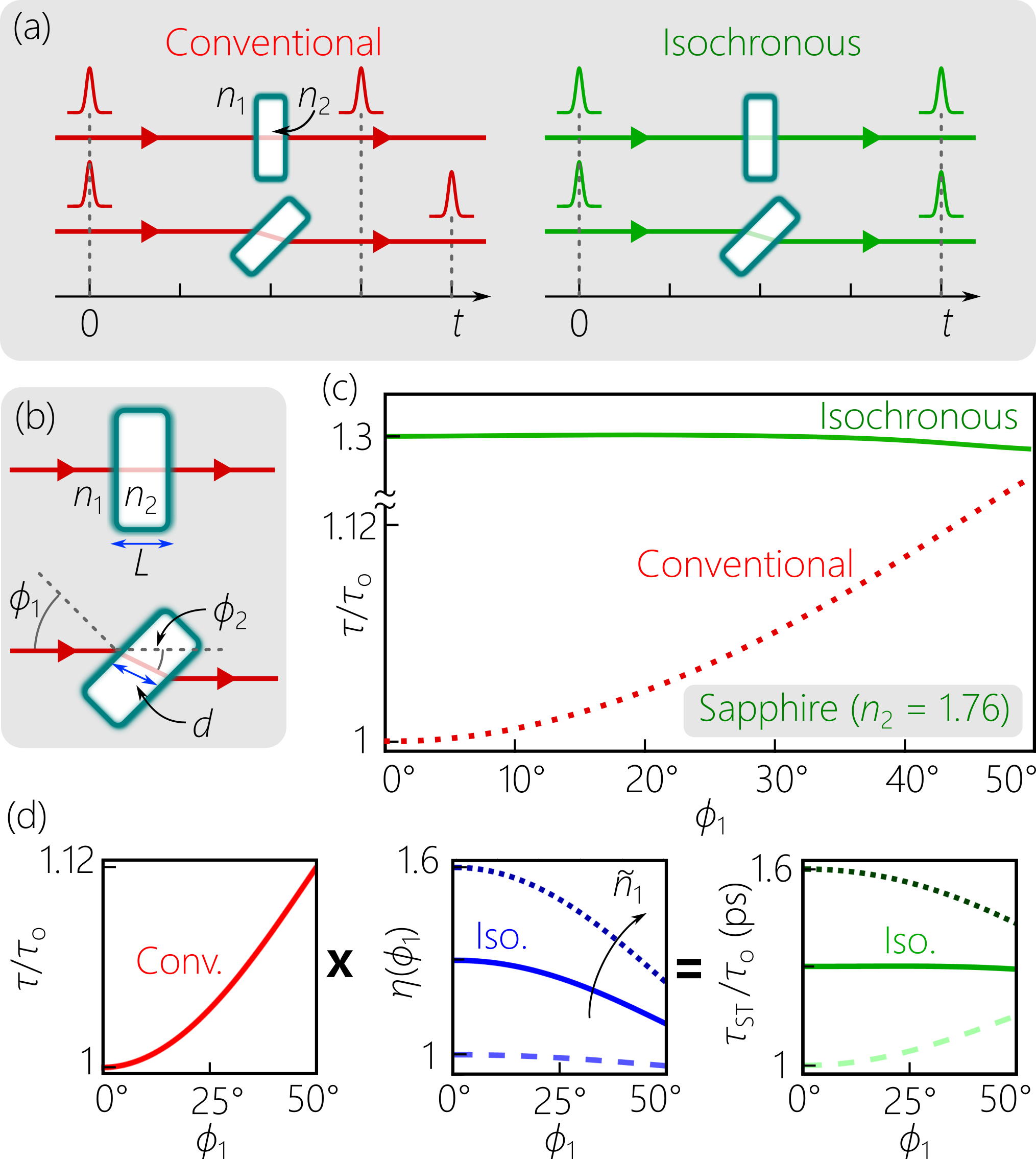}
  \end{center}
  \caption{a) Concept of isochronous ST wave packets. A conventional wave packet incurs a larger group delay traversing a layer at oblique incidence than at normal incidence. In contrast, an isochronous ST wave packet maintains the same group delay independently of the incident angle. (b) The path length $d$ across the layer at oblique incidence increases with the incident angle $\phi_{1}$. (c) Calculated oblique-incidence delay $\tau$ across a sapphire layer normalized to the normal-incidence delay $\tau_{\mathrm{o}}$. (d) The delay $\tau_{\mathrm{ST}}$ for a ST wave packet is the product of the delay $\tau$ for a conventional wave packet and a factor $\eta(\phi_{1})$ dictated by the group index $\widetilde{n}_{1}$. At the isochronous condition, $\tau_{\mathrm{ST}}$ is independent of $\phi_{1}$.}
    \label{Fig:Slab}
\end{figure}

We show here that the refraction of a recently developed family of pulsed beams denoted `space-time' (ST) wave packets \cite{Kondakci16OE,Parker16OE,Kondakci17NP,Yessenov19OPN} offers a unique configuration, whereby the group delay incurred by the refracted wave packet traversing a homogeneous, isotropic dielectric slab remains fixed with increasing incident angle despite the increase in path length. We thus uncover a new optical refractory phenomenon: the existence of \textit{isochronous} ST wave packets. This counter-intuitive phenomenon stems from the unusual behavior of ST wave packets when refracting across a planar interface, as verified recently in \cite{Bhaduri20NP}. Specifically, the group velocity of certain ST wave packets increases with increasing incident angle. The isochronous condition is reached when the increase in the refracted group velocity at any incident angle compensates for the larger associated path length in the slab. Consequently, the wave packet emerges from the slab at a fixed delay over a large span of incident angles. We confirm these predictions experimentally in three materials over a wide range of refractive indices: MgF$_2$ ($n\!\approx\!1.38$), sapphire ($n\!\approx\!1.76$), and ZnSe ($n\!\approx\!2.51$). The isochronous condition separates two regimes, one in which the delay increases with angle of incidence as expected, and one in which it anomalously \textit{decreases}. This rare example of an optical isochronous configuration is a critical experimental step towards realizing blind synchronization of remote clocks as proposed in \cite{Bhaduri20NP}.

Consider the scenario depicted in Fig.~\ref{Fig:Slab}(a) where a wave packet is incident onto a planar layer of a \textit{non-dispersive} material of refractive index $n_{2}$ and thickness $L$ at an angle $\phi_{1}$ with respect to the layer normal from a medium of refractive index $n_{1}$. At normal incidence, the group delay upon traversing the layer is $\tau_{\mathrm{o}}\!=\!n_{2}L/c$, where $c$ is the speed of light in vacuum. At oblique incidence the group delay increases monotonically with $\phi_{1}$, $\tau(\phi_{1})\!=\!\tau_{\mathrm{o}}/\cos{\phi_{2}}$, because of the larger path length $d\!>\!L$ [Fig.~\ref{Fig:Slab}(b)]; where $n_{1}\sin{\phi_{1}}\!=\!n_{2}\sin{\phi_{2}}$. We pose the following question: is it possible to synthesize an isochronous optical wave packet, one for which the group delay at oblique incidence $\tau(\phi_{1})$ is \textit{independent} of $\phi_{1}$, so that the delay incurred is fixed despite the longer propagation distance [Fig~\ref{Fig:Slab}(a)]? Realizing this isochronous condition requires that the group velocity $\widetilde{v}_{2}$ in the layer increases with $\phi_{1}$, $\widetilde{v}_{2}(\phi_{1})\!\propto\!1/\cos{\phi_{1}}$, to counterbalance the longer propagation distance $d(\phi_{1})\!\propto\!\cos{\phi_{2}}$ [Fig~\ref{Fig:Slab}(b)], so that $\tau(\phi_{1})\!=\!d(\phi_{1})/\widetilde{v}_{2}(\phi_{1})$ is constant [Fig~\ref{Fig:Slab}(c)]. For conventional wave packets, oblique incidence does \textit{not} impact $\widetilde{v}_{2}$. However, it was recently shown that refraction of ST wave packets follows an unusual rule whereby $\widetilde{v}_{2}$ depends not only on $n_{2}$, but also on $n_{1}$, the group velocity of the incident wave packet $\widetilde{v}_{1}$, \textit{and} the angle of incidence $\phi_{1}$. This unique behavior offers the possibility of an isochronous configuration.

\begin{figure}[t!]
  \begin{center}
  \includegraphics[width=\linewidth]{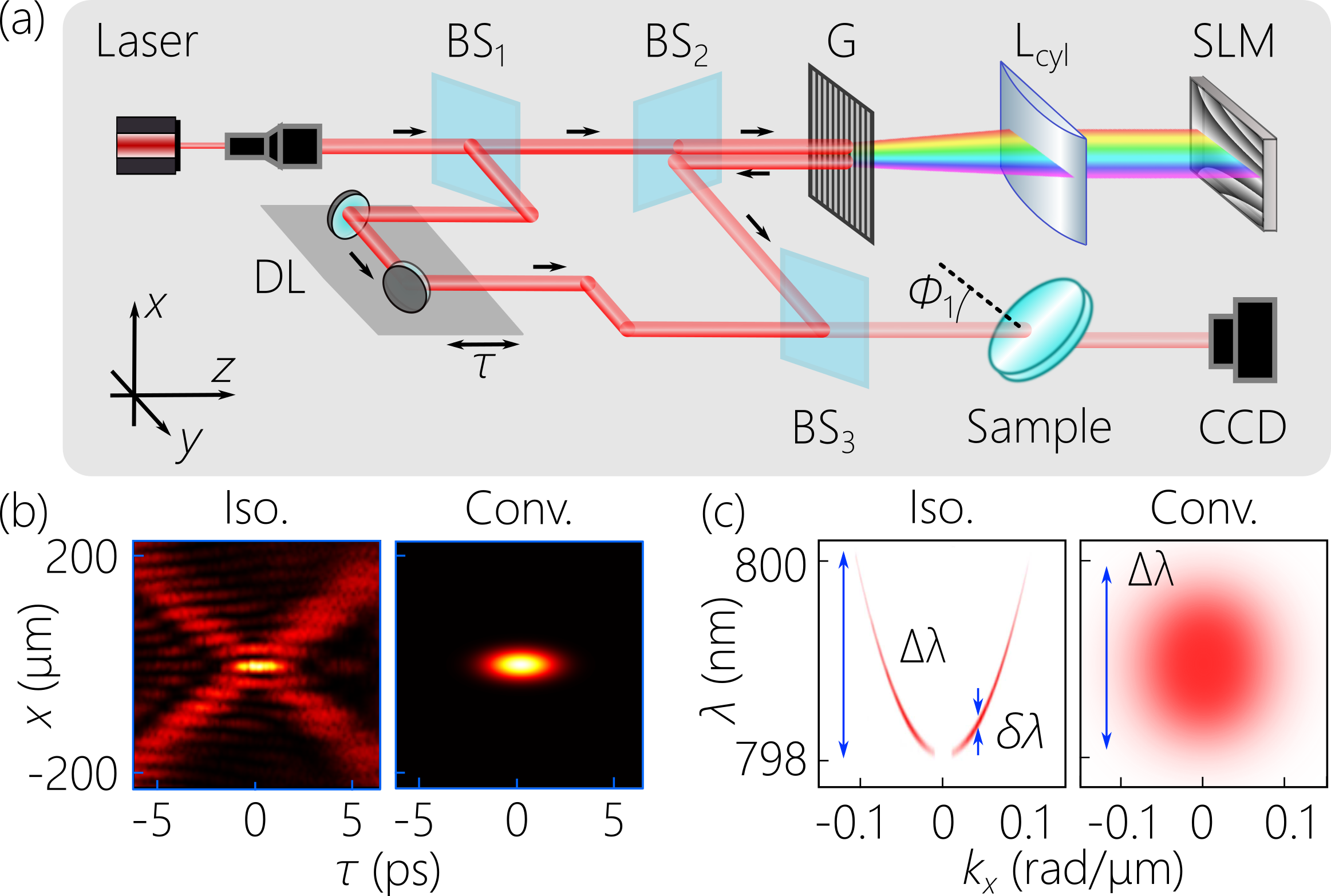}
  \end{center}
  \caption{(a) Schematic of the setup for synthesizing and characterizing isochronous ST wave packets. BS: Beam splitter, G: grating, L$_{\mathrm{cyl}}$: cylindrical lens, SLM: spatial light modulator, DL: delay line. (b) The spatio-temporal intensity profiles for conventional and isochronous ST wave packets (the latter is measured) and (c) their corresponding spatio-temporal spectrum.}
    \label{Fig:Schematic}
\end{figure}

The key distinguishing characteristic of ST wave packets is that each spatial frequency is associated with a single wavelength \cite{Donnelly93ProcRSLA,Saari04PRE,Longhi04OE,Yessenov19PRA,Yessenov19OE} so as to realize non-differentiable angular dispersion \cite{Hall21arxiv}, which enables realizing wave packets endowed with propagation invariance \cite{Kiselev07OS,Turunen10PO,FigueroaBook14} and tunable group velocities \cite{Salo01JOA,Kondakci19NC,Bhaduri19Optica}. Specifically, the spatio-temporal spectrum of a ST wave packet in a non-dispersive medium of index $n$ lies at the intersection of the light cone $k_{x}^{2}+k_{z}^{2}\!=\!n^{2}(\tfrac{\omega}{c})^{2}$ with a spectral plane $\mathcal{P}(\theta)$ defined by the equation $\omega\!=\!\omega_{\mathrm{o}}+(k_{z}-nk_{\mathrm{o}})c\tan{\theta}$, which is parallel to the $k_{x}$-axis and makes an angle $\theta$ (the spectral tilt angle) with respect to the $k_{z}$-axis \cite{Donnelly93ProcRSLA,Kondakci17NP}. Here $x$ and $z$ are the transverse and axial coordinates, respectively; $k_{x}$ and $k_{z}$ are the corresponding components of the wave vector; $\omega$ is the angular frequency; and $\omega_{\mathrm{o}}$ is a fixed frequency whose corresponding free-space wave number is $k_{\mathrm{o}}\!=\!\tfrac{\omega_{\mathrm{o}}}{c}$. For simplicity, but without loss of generality, we hold the field uniform along the transverse $y$-coordinate ($k_{y}\!=\!0$). This configuration imposes a precise association between each $k_{x}$ with a single $\omega$. The ST wave packet is \textit{propagation invariant}: it is transposed rigidly in the medium without diffraction or dispersion \cite{Kondakci17NP,Porras17OL,Efremidis17OL,Wong17ACSP2} at a group velocity $\widetilde{v}\!=\!c\tan{\theta}\!=\!c/\widetilde{n}$ ($\widetilde{n}\!=\!c/\widetilde{v}\!=\!\cot{\theta}$ is the group index), which is determined by the wave-packet spatio-temporal spectral structure \cite{Yessenov19PRA}. The ST wave packet is subluminal when $\cot{\theta}\!>\!n$ and superluminal when $\cot{\theta}\!<\!n$. Assuming paraxial ($\Delta k_{x}\!\ll\!k_{\mathrm{o}}$) and narrowband ($\Delta\omega\!\ll\!\omega_{\mathrm{o}}$) conditions, the conic section at the intersection of the light-cone with $\mathcal{P}(\theta)$ in the vicinity of $k_{x}\!=\!0$ can be approximated by a parabola:
\begin{equation}\label{Eq:Parabola}
\frac{\omega}{\omega_{\mathrm{o}}}=1+\frac{1}{n(n-\widetilde{n})}\;\frac{k_{x}^{2}}{2k_{\mathrm{o}}^{2}}.
\end{equation}
This association between the spatial and temporal frequencies ($k_{x}$ and $\omega$, respectively) can be achieved by the pulse shaper developed in \cite{Kondakci17NP,Bhaduri19Optica}; see Fig.~\ref{Fig:Schematic}(a) below.

\begin{figure}[t!]
  \begin{center}
  \includegraphics[width=\linewidth]{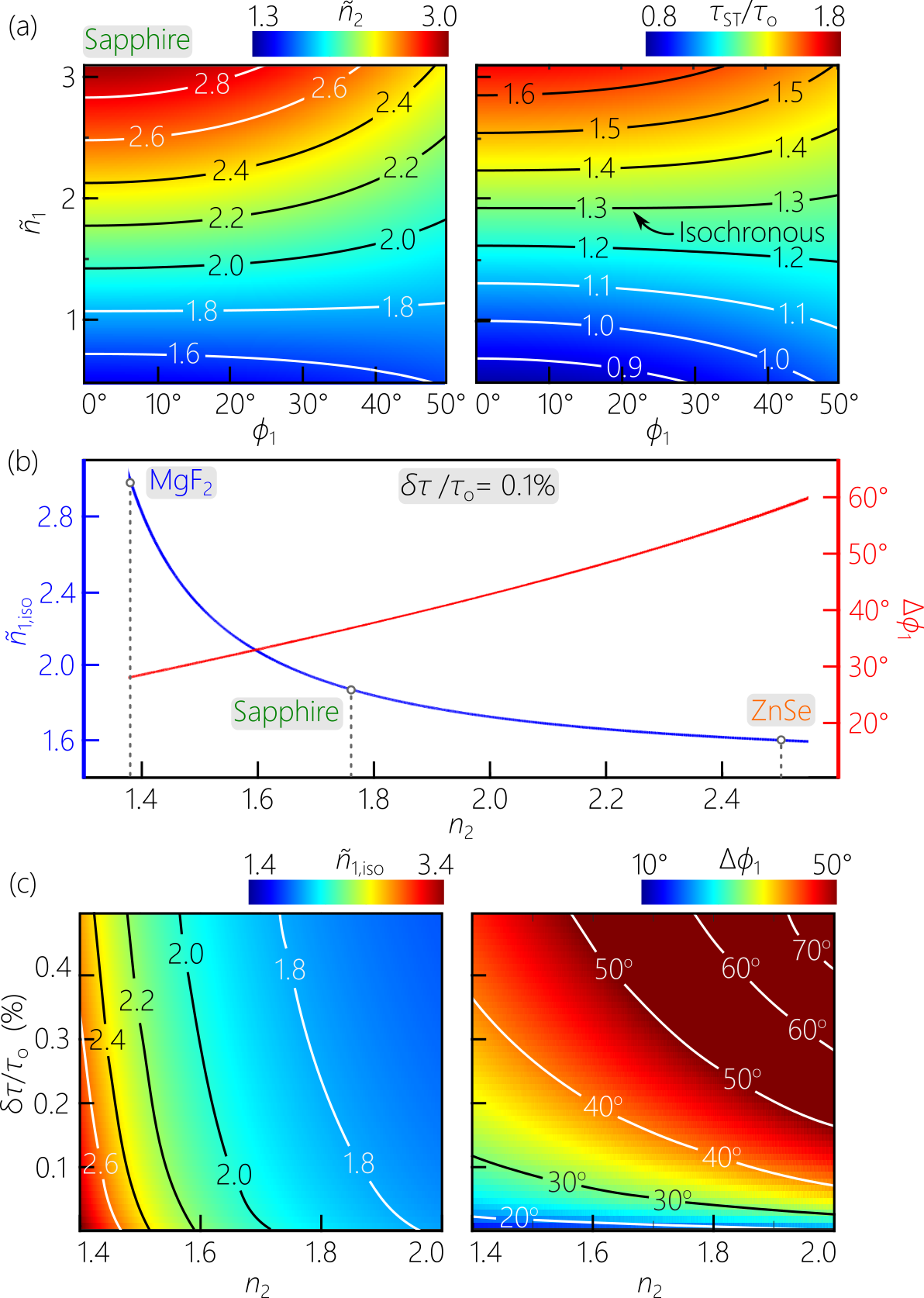}
  \end{center}
  \caption{(a) Calculated group index $\widetilde{n}_{2}$ of the refracted ST wave packet and the normalized delay $\tau_{\mathrm{ST}}/\tau_{\mathrm{o}}$ (Eq.~\ref{Eq:GroupDelay}) as a function of the incident wave packet group index $\widetilde{n}_{1}$ and the incident angle $\phi_{1}$, the latter corresponding to Fig.~\ref{Fig:Slab}(c). The calculations are performed for a planar layer of sapphire $n_{2}\!=\!1.76$ assuming incidence from free space $n_{1}\!=\!1$. (b) Calculated isochronous condition $\widetilde{n}_{1}^{(\mathrm{iso})}$ and the associated angular acceptance range $\Delta\phi_{1}$ for different materials, assuming incidence from free space $n_{1}\!=\!1$ and a delay tolerance of $\delta\tau/\tau_{\mathrm{o}}\!=\!0.1\%$. The three materials used in our experiments (MgF$_2$, sapphire, and ZnSe) are identified. (c) Calculated $\widetilde{n}_{1}^{(\mathrm{iso})}$ and $\Delta\phi_{1}$ for different materials while varying the delay tolerance $\delta\tau/\tau_{\mathrm{o}}$.}
    \label{Fig:Calculation}
\end{figure}

The refraction of a ST wave packet across a planar interface between two transparent, isotropic, homogeneous, non-dispersive dielectrics of refractive indices $n_{1}$ and $n_{2}$ is governed by \cite{Bhaduri20NP}:
\begin{equation}\label{Eq:Law}
n_{1}(n_{1}-\widetilde{n}_{1})\cos^{2}{\phi_{1}}=n_{2}(n_{2}-\widetilde{n}_{2})\cos^{2}{\phi_{2}}.
\end{equation}
where $\widetilde{n}_{1}$ and $\widetilde{n}_{2}$ are the group indices for the incident and transmitted wave packets, respectively, and $\phi_{1}$ and $\phi_{2}$ are their angles with respect to the normal to the interface. That is, $\widetilde{n}_{2}$ depends on the refractive indices of \textit{both} media, and also on $\widetilde{n}_{1}$ and $\phi_{1}$. This relationship reflects the invariance of the so-called `spectral curvature' of the ST wave packet, $n(n-\widetilde{n})\cos^{2}{\phi}$. This new optical invariant results from combining the conservation of transverse momentum and energy ($k_{x}$ and $\omega$) across a planar interface for ST wave packets in which the spatial and temporal degrees of freedom are inextricably linked \cite{Bhaduri20NP}.

If the second medium is a slab of thickness $L$, then the group delay incurred by the ST wave packet traversing it is:
\begin{equation}\label{Eq:GroupDelay}
\tau_{\mathrm{ST}}(\phi_{1})=\widetilde{n}_{2}\frac{d}{c}=\frac{\tau_{\mathrm{o}}}{\cos{\phi_{2}}}\left\{1+\frac{n_{1}^{2}\cos^{2}{\phi_{1}}}{n_{2}^{2}\cos^{2}{\phi_{2}}}\left(\frac{\widetilde{n}_{1}}{n_{1}}-1\right)\right\},
\end{equation}
where $\tau_{\mathrm{o}}\!=\!n_{2}\ell/c$ is the group delay for a conventional wave packet at normal incidence. Intuitively we expect $\tau_{\mathrm{ST}}(\phi_{1})$ to increase with $\phi_{1}$ because of the increase in the traveled distance. This monotonic trend is however counterbalanced by the term in the parentheses that decreases with $\phi_{1}$ in the subluminal regime $\widetilde{n}_{1}\!>\!n_{1}$, whereupon the group velocity of the refracted ST wave packet increases with $\phi_{1}$ \cite{Bhaduri20NP}.

We plot in Fig.~\ref{Fig:Calculation}(a) the group index $\widetilde{n}_{2}$ of the transmitted wave packet traversing a planar interface from free space ($n_{1}\!=\!1$) to sapphire ($n_{2}\!=\!1.76$ at a wavelength of $\lambda\!=\!800$~nm) while varying the group index $\widetilde{n}_{1}$ of the incident wave packet and its incident angle $\phi_{1}$. When $\widetilde{n}_{1}\!=\!1$ (a plane-wave pulse \cite{Yessenov19PRA}), $\widetilde{n}_{2}\!=\!n_{2}$ independently of $\phi_{1}$ as usual, and the group delay $\tau\!=\!\widetilde{n}_{2}d/c$ increases monotonically with $\phi_{1}$ because of the increased path length across the slab. For all other values of $\widetilde{n}_{1}$, the refracted group index $\widetilde{n}_{2}$ varies with $\phi_{1}$. We require that $\widetilde{n}_{2}$ \textit{decrease} with $\phi_{1}$ to counterbalance the increased path length. It is clear in Fig.~\ref{Fig:Calculation}(a), that this condition occurs when $\widetilde{n}_{1}\!>\!n_{1}$; i.e., in the subluminal regime. When we plot $\tau_{\mathrm{ST}}$ normalized with respect to $\tau_{\mathrm{o}}$ in Fig.~\ref{Fig:Calculation}(a), we observe that a flat iso-delay contour occurs at a particular value of $\widetilde{n}_{1}\!=\!\widetilde{n}_{1,\mathrm{iso}}$. This ST wave packet encounters a constant delay as $\phi_{1}$ varies over a broad span. 

We plot in Fig.~\ref{Fig:Calculation}(b) the calculated group index $\widetilde{n}_{1,\mathrm{iso}}$ for the incident ST wave packet that satisfies the isochronous condition in a medium of index $n_{2}$ (assuming incidence from free space $n_{1}\!=\!1$). We define this condition as follows: for a given $n_{2}$, we calculate the normalized group delay $\tau_{\mathrm{ST}}(\phi_{1};\widetilde{n}_{1})/\tau_{\mathrm{o}}$. For each value of $\widetilde{n}_{1}$, we determine the maximum incident angle $\phi_{1}$ that maintains the delay $\tau_{\mathrm{ST}}(\phi_{1};\widetilde{n}_{1})$ within the range $\tau_{\mathrm{ST}}(0;\widetilde{n}_{1})\pm\delta\tau$. The group index $\widetilde{n}_{1}$ for the incident wave packet that yields the maximum incident angle within this range for the delay is denoted $\widetilde{n}_{1,\mathrm{iso}}$. We plot the calculated $\widetilde{n}_{1,\mathrm{iso}}$ for $\delta\tau/\tau_{\mathrm{o}}\!=\!0.1\%$ and the corresponding isochronous angular range $\Delta\phi_{1}$. Of course, the value of $\widetilde{n}_{1,\mathrm{iso}}$ and $\Delta\phi_{1}$ depend on the choice of $\delta\tau$. We plot both $\widetilde{n}_{1,\mathrm{iso}}$ and $\Delta\phi_{1}$ as we vary the delay tolerance $\delta\tau/\tau_{\mathrm{o}}$ in Fig.~\ref{Fig:Calculation}(c) for different materials. It is clear that relaxing the delay tolerance increases the angular range $\Delta\phi_{1}$ over which the isochronous condition is maintained. Furthermore, relaxing $\delta\tau$ reduces $\widetilde{n}_{1,\mathrm{iso}}$, which approaches $n_{1}\!=\!1$. This makes synthesizing the isochronous ST wave packet more convenient \cite{Yessenov19PRA}.

To measure the group delay $\tau_{\mathrm{ST}}(\phi_{1})$ for planar samples, we make use of the interferometric arrangement in Fig.~\ref{Fig:Schematic}(a), which is based on our previous work in \cite{Kondakci19NC,Bhaduri19Optica,Bhaduri20NP}. We employ 100-fs pulses from a mode-locked Ti:sapphire laser at a central wavelength of 800~nm directed to a two-path interferometer. In one path, we place the setup for synthesizing the ST wave packets in which the spectrum of the laser pulses is resolved with a diffraction grating (1200~lines/mm), and the first diffraction order is collimated by a cylindrical lens (focal length 50~cm) and directed to a reflective phase-only spatial light modulator (SLM; Hamamatsu X10468-02) that imparts a two-dimensional phase distribution to the spectrally resolved wave front. This phase distribution is designed to assign a specific spatial frequency $k_{x}$ to each wavelength $\lambda$ according to the constraint in Eq.~\ref{Eq:Parabola}. Tuning the spectral tilt angle $\theta_{1}$ of the ST wave packet to change its group index $\widetilde{n}_{1}\!=\!\cot{\theta_{1}}$ is achieved by sculpting the SLM phase \cite{Yessenov19PRA}. A delay line $\tau$ is placed in the path of the reference arm of the interferometer. The wave packet shaper spectrally filters the initial pulses to a bandwidth of $\Delta\lambda\!\approx\!0.5$~nm, such that the narrower reference pulses are a valid probe to reconstruct the spatio-temporal profile of the ST wave packet by sweeping $\tau$ [Fig.~\ref{Fig:Schematic}(b)]. The group index $\widetilde{n}_{1}$ of the synthesized ST wave packet is verified in two ways. First, the differential group delay between the ST wave packet (traveling at $c/\widetilde{n}_{1}$) and the reference pulse (traveling at $c$) is measured after placing the detector at two different axial positions, from which we can estimate $\widetilde{n}_{1}$. Second, we measure the spatio-temporal spectrum in the $(k_{x},\lambda)$-plane [Fig.~\ref{Fig:Schematic}(c)] using a combination of grating and Fourier-transforming lens (not shown in the setup for simplicity) \cite{Kondakci19NC}.

\begin{figure}[t!]
  \begin{center}
  \includegraphics[width=\linewidth]{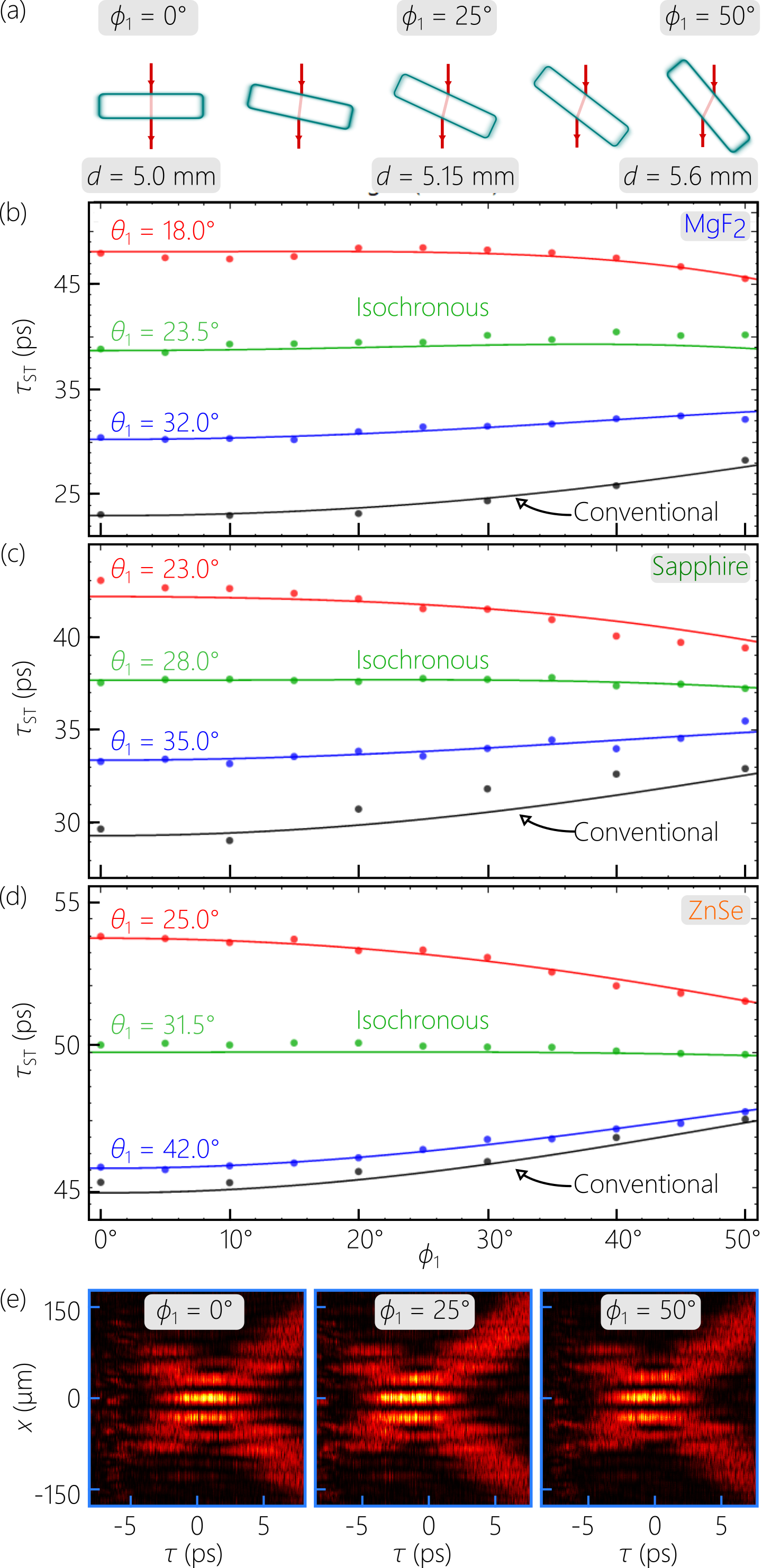}
  \end{center}
  \caption{(a) Illustration of the path across a planar layer with $\phi_{1}$. (b) Measured group delay $\tau$ with $\phi_{1}$ for MgF$_2$. The three curves correspond to different spectral tilt angles $\theta_{1}$ for the incident wave packet from free space, compared to a conventional pulse. The dots are data points. (c) Same as (b) for sapphire. (d) Same as (b) for ZnSe. A small modification is made to Eq.~\ref{Eq:Law} in this case to account for dispersion in ZnSe. (e) Measured spatio-temporal profiles $I(x,\tau)$ for the isochronous wave packets after traversing the Sapphire layer at $\phi_{1}\!=\!0^{\circ}$, $25^{\circ}$, and $50^{\circ}$.}
    \label{Fig:Measurements}
\end{figure}

The isochronous condition is verified in planar samples of MgF$_2$, sapphire, and ZnSe all of thickness $L\!=\!5$~mm, which are placed in the common path of the ST wave packet and reference pulse after the interferometer, as shown in Fig.~\ref{Fig:Schematic}(a). Each sample is rotated about the $y$-axis in increments of $5^{\circ}$, resulting in an increase in path length from 5~mm to $\approx\!6$~mm in MgF$_2$, 5.6~mm in sapphire, and 5.25~mm in ZnSe. The group delay $\tau_{\mathrm{ST}}(\phi_{1})$ is measured for each incident angle $\phi_{1}$, and the results for the three layers are plotted in Fig.~\ref{Fig:Measurements}(b-d). In each case, we plot results for three group indices of the incident ST wave packets: $\widetilde{n}_{1}\!=\!\widetilde{n}_{1,\mathrm{iso}}$ where $\tau_{\mathrm{ST}}(\phi_{1})$ is independent of $\phi_{1}$ over an extended range of incident angles, $\widetilde{n}_{1}\!<\!\widetilde{n}_{1,\mathrm{iso}}$ where $\tau_{\mathrm{ST}}(\phi_{1})$ increases with $\phi_{1}$ as expected, and $\widetilde{n}_{1}\!>\!\widetilde{n}_{1,\mathrm{iso}}$ where $\tau_{\mathrm{ST}}(\phi_{1})$ anomalously drops with $\phi_{1}$. We compare the results in each case to those for a conventional wave packet (plane-wave pulse) corresponding to $\widetilde{n}_{1}\!=\!1$ ($\theta\!=\!45^{\circ}$) obtained by idling the SLM in Fig.~\ref{Fig:Schematic}(a). The case of ZnSe is particularly striking where the isochronous condition is maintained over the largest span of incident angles despite the changes in the delay $\tau_{\mathrm{ST}}(\phi_{1})$ with $\phi_{1}$.

We hypothesized in \cite{Bhaduri20NP} that the appropriate ST wave packets can be utilized to blindly synchronize a transmitter with multiple remote receivers at different unknown locations at the same depth beyond an interface between two media. In such a configuration, the sum of the delays in the two media must be invariant with angle of incidence at the interface. Our results here regarding isochronous ST wave packets are a significant step towards such a goal by demonstrating the invariant delay in the second medium (the slab here) with angle of incidence. The maximum distance over which the ST wave packets can be used (the maximum thickness of the slab here) is limited by the so-called `spectral uncertainty' $\delta\lambda$, which is the unavoidable finite bandwidth associated with each spatial frequency \cite{Yessenov19OE}. In our experiments here, $\delta\lambda\!\sim\!25$~pm, as determined by the spectral resolution of the grating in Fig.~\ref{Fig:Schematic}(a). The maximum propagation distance exceeds the 5-mm-thickness of the slabs used here. This is confirmed in Fig.~\ref{Fig:Measurements}(e), where only minimal changes are observed in the spatio-temporal profile of the wave packet after traversing the sapphire slab at all values of $\phi_{1}$ of interest.

In conclusion, we have demonstrated for the first time, to the best of our knowledge, isochronous optical wave packets: pulsed beams that incur the same group delay after traversing a dielectric slab at any incident angle despite the different path lengths. In our realization, ST wave packets synthesized in free space with a particular group index can satisfy this condition and maintain an invariant group delay over a wide range of incident angles. Isochronous ST wave packets may have applications in clock-synchronization, in free-space optical communications, and in nonlinear optics. Finally, our work here is based on the refraction of propagation-invariant ST wave packet in non-dispersive media. It will be interesting to extend this work to dispersive materials \cite{Malaguti08OL,Malaguti09PRA}, and to study the refraction of recently developed ST wave packets that undergo controllable axial evolution, such as accelerating or decelerating wave packets \cite{Yessenov20PRL2}, and those endowed with axial spectral encoding \cite{Motz20arxiv} or group-velocity dispersion in free space \cite{Yessenov21arxiv}.

\section*{Funding}
U.S. Office of Naval Research (ONR) contract N00014-17-1-2458 and ONR MURI contract N00014-20-1-2789.

\vspace{2mm}
\noindent
\textbf{Disclosures.} The authors declare no conflicts of interest.

\bibliography{diffraction}
\end{document}